# Novel Superconducting Ternary Hydrides under High Pressure

Bangshuai Zhu[1], Dexi Shao[2], Cuiying Pei[1], Qi Wang[1,3], Juefei Wu[1*], Yanpeng Qi[1,3,4*]

1. School of Physical Science and Technology, ShanghaiTech University, Shanghai 201210, China

2. School of Physics, Hangzhou Normal University, Hangzhou 31121, China

3. ShanghaiTech Laboratory for Topological Physics, ShanghaiTech University, Shanghai 201210, China

4. Shanghai Key Laboratory of High-resolution Electron Microscopy, ShanghaiTech University, Shanghai 201210, China

* Correspondence should be addressed to Y.P.Q. (qiyp@shanghaitech.edu.cn) or J.W. (wujf@shanghaitech.edu.cn)

## ABSTRACT

The abundant chemical compositions in ternary hydrides bring much more possibility to explore high temperature superconductors under lower pressure. Here we constructed 115 ternary hydrides on the basis of the elements substitution using 16 metal elements within 5 reported prototype structures. We conducted a three-step approach to screen and study these candidate structures in the aspects of dynamical stability, formation energy and relative enthalpy, respectively. Based on this approach, we found three meta-stable compounds with hydrogen clathrate cages in the space group of $P\bar{3}m1$, including $Y_2CdH_{18}$, $Y_2InH_{18}$ and $Ca_2SnH_{18}$. All of the structures are superconductive under high pressure with $T_c$ above 110 K, which is larger than the superconductive temperature of liquid nitrogen. Our study enriches the database of novel ternary hydrides under high pressure, and provides insight for future theoretical and experimental researches.

## Introduction

Realizing room temperature superconductivity is one of the ultimate goals in the field of condensed matter physics. Inspired by the theory of "chemical precompression" [1], hydrogen-rich compounds under pressure have undergone extensive researches and became potential materials for the pursuit of room temperature superconductivity. In recent years, theoretical and experimental researches reported a series of superconducting binary hydrides with high critical transition temperatures ($T_c$) [2-10]. In particular, the covalent hydride $H_3S$ achieved a $T_c$ of 203 K at 155 GPa [2, 3, 11, 12], and the clathrate hydride $LaH_{10}$ has a $T_c$ of 250-260 K at 180-200 GPa [6, 13]. In addition, other binary hydrides such as $YH_9$ and $CaH_6$ also exhibit $T_c$ over 200 K [7, 10, 14]. Despite the stabilizing pressures for these binary hydrides are lower than hydrogen metallization pressure 500 GPa [15-17], the pressure above 150 GPa limit their practical applications. The following goal is to further reduce the pressure with the premise of high $T_c$. Therefore, it is worthwhile to study superconducting ternary hydrides under high pressure, owing to the more abundant chemical compositions that could provide more possibility to achieve this goal [18].

Compared with binary hydrides, the research on ternary hydrides is much more challenging. From the experimental aspect, the appropriate element selections and specific synthesis conditions require much experimental resources. In terms of theoretical simulations, though the theoretical simulations could greatly help reduce the experimental cost, the calculations on the whole ternary phase diagrams are very expensive and time-consuming. Thus, the theoretical studies of ternary hydrides need some certain strategies. Based on recent studies [19-36], we summarize two kinds of strategies to obtain unique ternary hydrides under high pressure. One of the strategy is combing two kinds of superconducting binary hydrides under high pressure [23-25, 28-31]. For example, Semenok *et al*. synthesized a series of La-Y-H compounds and the reported $LaYH_{20}$ has a $T_c$ of 253 K at 183 GPa [23]. Zhao *et al*. combined Ca-H [7, 37] and Y-H [6] and predicted $Y_3CaH_{24}$ [25] with $T_c$ of 242-258 K at 150 GPa. The other strategy is performing elements substitution based on structure designing [20, 26, 27, 32-36]. Zhang *et al.* identified a fluorite-type backbone structure donated as "$AXH_8$" ($Fm\bar{3}m$). After researching elements combinations, they reported $LaBeH_8$ with the $T_c$ of 185 K at 20 GPa [20]. Jiang *et al.* studied $AC_2H_8$ compounds (A=Na, K, Mg, Al, Ga) and predicted $MgC_2H_8$ with $T_c$ of 55 K at 40 GPa [27]. Du *et al.* predicted $KGaH_3$ with $T_c$ of 146 K at 10 GPa based on the perovskite structure of $KInH_3$ ($Fm\bar{3}m$) [36]. Both strategies could improve the efficiency of studying novel ternary hydrides, and the substitution strategy is capable of screening some unexpected elements that bring about more possibility in ternary hydrides, such as Be in $LaBeH_8$ [20] and Zr in $CaZrH_{12}$ [26].

Thus, in this study, we have identified 5 ternary hydrides structures with clathrate features ($A_2XH_{18}$ ($P\bar{3}m1$) [25], $AXH_{20}$ ($P4/mmm$) [25], $AC_2H_8$ ($Fm\bar{3}m$) [27], $A_3XH_{24}$ ($Fm\bar{3}m$) [25] and $LaXH_8$ ($Fm\bar{3}m$) [20]) as prototypes to follow the strategy of elements substitution to search novel superconductors. These structures exhibit high symmetry that are conductive to high $T_c$. As for the elements, we considered Mg, Sr, Sc, La, Cu, Ag, Au, Zn, Cd, Hg, Ga, In, Tl, Ge, Sn and Pb for substitution. Although some of these elements like Cd and In have less contribution than La, Y and Ca in binary hydrides, they may open up new possibilities in ternary hydrides. We constructed 115 new structures and proposed a three-step approach to screen and evaluate candidate structures in the aspects of dynamical stability, formation energy and relative enthalpy, respectively. We have identified 3 candidate compounds including $Y_2CdH_{18}$,$Y_2InH_{18}$ and $Ca_2SnH_{18}$ within $A_2XH_{18}$ ($P\bar{3}m1$). These predicted structures are meta-stable and have potential for synthesizing under high pressure. Then we calculated their electronic structures and superconducting properties. The estimated $T_c$ values of the three structures are higher than 110 K, which is above the liquid nitrogen temperature 77 K.

## Computation Methods

Structure optimization and electronic structures calculations were performed by utilizing the Vienna *ab initio* Simulation Package (VASP) [38] within the framework of density functional theory (DFT) [39, 40]. The Perdew-Burke-Ernzerhof (PBE) functional based on the generalized gradient approximation (GGA) was chosen to describe the exchange-correlation interactions [41, 42], and the projector augmented wave (PAW) method [43] was adopted with a valence configuration of 1*s* for H, 4*s*4*p*5*s*4*d* for Y, 3*s*3*p*4*s* for Ca, 5*s*4*d* for Cd, and 5*s*5*p* for In and Sn. We set the plane-wave kinetic-energy cutoff to 450 eV, and the Brillouin zone was sampled by the Monkhorst-Pack [44] scheme of 2π×0.03 $Å^{-1}$. The convergence criteria were $10^{-6}$ eV for energy and 0.003 eV/Å for atomic forces, respectively. Phonon spectrum calculations were performed by utilizing the supercell finite displacement method implemented in the PHONOPY package [45], 2×2×2 supercells were applied to all structures. To study the chemical bonding properties, crystal orbital Hamilton populations (COHP) analysis was performed by using the LOBSTER 4.1.0 package with the pbeVaspFit2015 basis set [46].

The EPC coefficients $\lambda$ were calculated by the QUANTUM ESPRESSO (QE) package [47] using density-functional perturbation theory (DFPT) [48]. We selected the ultrasoft pseudopotential with a kinetic energy cutoff of 100 Ry for $Y_2InH_{18}$ and $Ca_2SnH_{18}$, while 90 Ry for $Y_2CdH_{18}$, a 3×3×4 q-grid were chosen for the $P\bar{3}m1$ phase, the phonon spectra are rechecked by QE. The Allen-Dynes modified McMillan equation [49] and the self-consistent solution of the Eliashberg equation [50] were used to estimate the superconducting transition temperature $T_c$.

$$T_c = \frac{\omega_{\log}}{1.2}\exp\left(\frac{-1.04(1+\lambda)}{\lambda-\mu^*(1+0.62\lambda)}\right) \tag{1}$$

The $\mu^*$ is Coulomb pseudopotential and we used the value 0.10 and 0.13. $\omega_{\log}$ is the logarithmic average frequency, and $\lambda$ is the electron-phonon coupling coefficient, as difined as:

$$\lambda = 2\int_0^{\omega}\frac{\alpha^2F(\omega)}{\omega}d\omega \tag{2}$$

$$\omega_{log} = \exp\left[\frac{2}{\lambda}\int_0^{\omega}\frac{\alpha^2F(\omega)}{\omega}\log\omega\, d\omega\right] \tag{3}$$

## Results and discussion

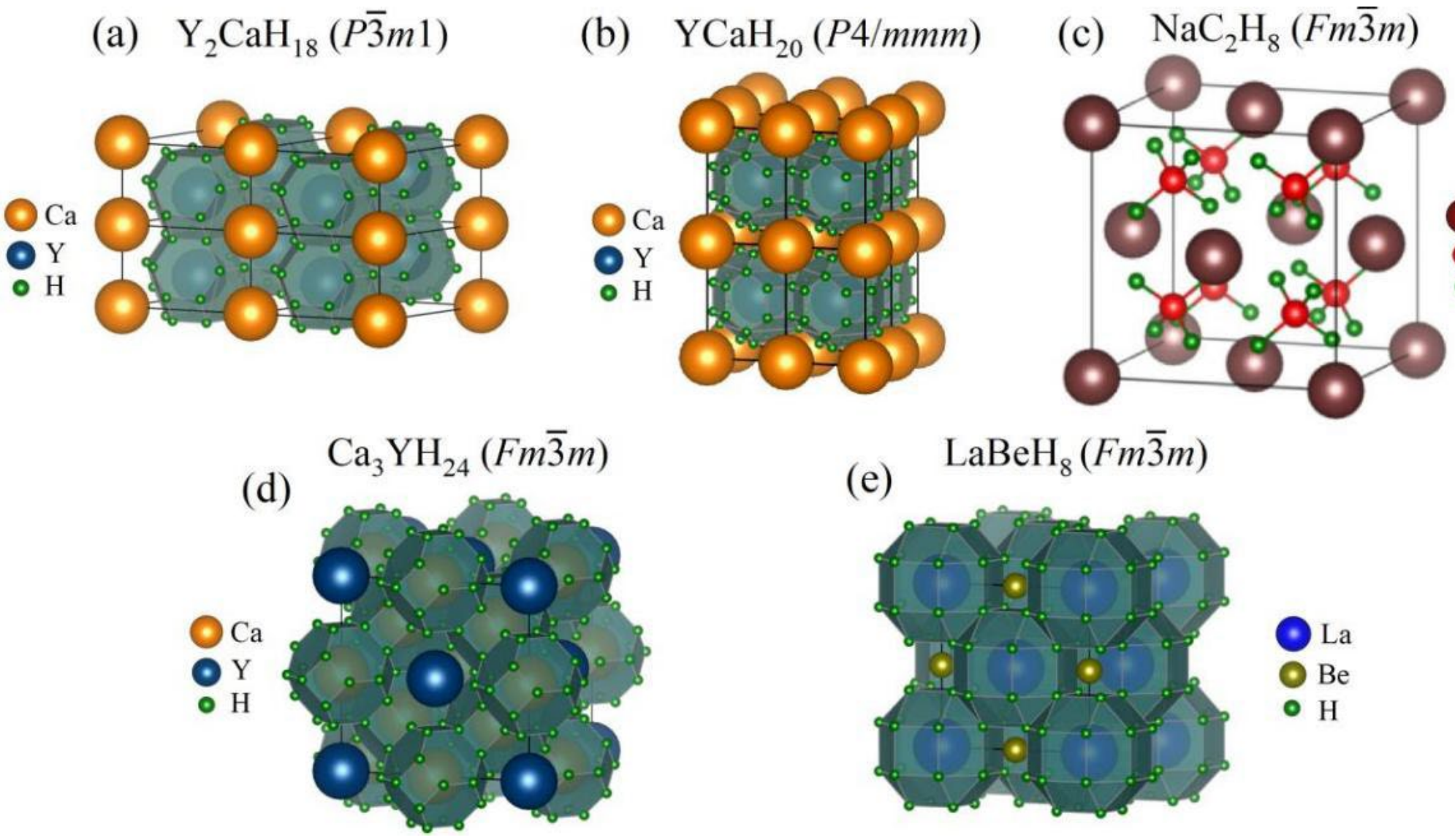

FIG. 1. The crystal structures of (a) $Y_2CaH_{18}$ ($P\bar{3}m1$), (b) $YCaH_{20}$ ($P4/mmm$), (c) $NaC_2H_8$ ($Fm\bar{3}m$), (d) $Ca_3YH_{24}$ ($Fm\bar{3}m$) and (e) $LaBeH_8$ ($Fm\bar{3}m$) [20, 25, 27].

In this study, we chose five ternary hydrides as prototype structures. Fig. 1 illustrates the structures of $Y_2CaH_{18}$ ($P\bar{3}m1$), $YCaH_{20}$ ($P4/mmm$), $NaC_2H_8$ ($Fm\bar{3}m$), $Ca_3YH_{24}$ ($Fm\bar{3}m$) and $LaBeH_8$ ($Fm\bar{3}m$). Among them, the hydrogen atoms in $Y_2CaH_{18}$, $YCaH_{20}$, $Ca_3YH_{24}$ and $LaBeH_8$ form the interconnected frameworks of the hydrogen cages and surround different guest atoms, while the cubic $NaC_2H_8$ contains a *fcc* framework composed of Na atoms with light $CH_4$ molecules embedded. Based on these structures, we selected Mg, Sr, Sc, La, Cu, Ag, Au, Zn, Cd, Hg, Ga, In, Tl, Ge, Sn, Pb as the substitution elements and constructed 115 candidate structures. Then, we adopted a three-step approach to screen and evaluate these structures.

TABLE 1. The calculated phonon spectra of $A_2XH_{18}$ ($P\bar{3}m1$) and $AXH_{20}$ ($P4/mmm$) at 200 GPa. The A and X refer to different metal elements. A = (Ca, Y, Sr, Sc, Mg, La), X = (Cu, Ag, Au, Zn, Cd, Hg, Ga, In, Tl, Ge, Sn, Pb)

| Dynamical stability of $\mathbf{A_2XH_{18}}$ ($P\bar{3}m1$) at 200 GPa | | | | | |
|---|---|---|---|---|---|
| $Ca_2CuH_{18}$ | - | $Y_2AuH_{18}$ | - | $Sr_2SnH_{18}$ | - |
| $Ca_2AgH_{18}$ | - | $Y_2ZnH_{18}$ | - | $Sr_2PbH_{18}$ | - |
| $Ca_2AuH_{18}$ | - | $\mathbf{Y_2CdH_{18}}$ | - | $Sc_2CdH_{18}$ | - |
| $Ca_2ZnH_{18}$ | - | $Y_2HgH_{18}$ | - | $Sc_2GaH_{18}$ | - |
| $Ca_2CdH_{18}$ | - | $Y_2GaH_{18}$ | - | $Sc_2InH_{18}$ | - |
| $Ca_2HgH_{18}$ | - | $\mathbf{Y_2InH_{18}}$ | - | $Sc_2GeH_{18}$ | - |
| $Ca_2GaH_{18}$ | - | $Y_2TlH_{18}$ | - | $Sc_2SnH_{18}$ | - |
| $Ca_2InH_{18}$ | - | $Y_2GeH_{18}$ | - | $Sc_2PbH_{18}$ | - |
| $Ca_2TlH_{18}$ | - | $Y_2SnH_{18}$ | - | $Mg_2CdH_{18}$ | - |
| $Ca_2GeH_{18}$ | - | $Y_2PbH_{18}$ | - | $Mg_2InH_{18}$ | - |
| **$Ca_2SnH_{18}$** | + | $Sr_2CdH_{18}$ | - | $Mg_2SnH_{18}$ | - |
| $Ca_2PbH_{18}$ | - | $Sr_2GaH_{18}$ | - | $La_2CdH_{18}$ | - |
| $Y_2CuH_{18}$ | - | $Sr_2InH_{18}$ | - | $La_2InH_{18}$ | - |
| $Y_2AgH_{18}$ | - | $Sr_2GeH_{18}$ | - | $La_2SnH_{18}$ | - |
| Dynamical stability of **$AXH_{20}$** (*P*4/*mmm*) at 200 GPa | | | | | |
| $CaCuH_{20}$ | - | $CaTlH_{20}$ | - | $YCdH_{20}$ | - |
| $CaAgH_{20}$ | - | $CaGeH_{20}$ | - | $YHgH_{20}$ | - |
| $CaAuH_{20}$ | - | $CaSnH_{20}$ | - | $YGaH_{20}$ | - |
| $CaZnH_{20}$ | - | $CaPbH_{20}$ | - | $YInH_{20}$ | - |
| $CaCdH_{20}$ | - | $YCuH_{20}$ | - | $YTlH_{20}$ | - |
| $CaHgH_{20}$ | - | $YAgH_{20}$ | - | $YGeH_{20}$ | - |
| $CaGaH_{20}$ | - | $YAuH_{20}$ | - | $YSnH_{20}$ | - |
| $CaInH_{20}$ | - | $YZnH_{20}$ | - | $YPbH_{20}$ | - |

TABLE 2. The calculated phonon spectra of $MC_2H_8$ ($Fm\bar{3}m$) and $A_3XH_{24}$ ($Fm\bar{3}m$) and $LaXH_8$ ($Fm\bar{3}m$) at 200 GPa. A = (Ca, Y, Mg, La, Sc), X = (Cu, Ag, Au, Zn, Cd, Hg, Ga, In, Tl, Ge, Sn, Pb)

| Dynamical stability of A**$C_2H_8$** ($Fm\bar{3}m$) at 200 GPa | | | | | |
|---|---|---|---|---|---|
| $CuC_2H_8$ | **-** | **$ZnC_2H_8$** | + | **$InC_2H_8$** | + |
| **$AgC_2H_8$** | + | **$CdC_2H_8$** | + | **$TlC_2H_8$** | + |
| **$AuC_2H_8$** | + | **$HgC_2H_8$** | + | | |
| Dynamical stability of **$A_3XH_{24}$** ($Fm\bar{3}m$) at 200 GPa | | | | | |
| $Ca_3CuH_{24}$ | - | $Ca_3SnH_{24}$ | - | $Y_3TlH_{24}$ | - |
| $Ca_3AgH_{24}$ | - | $Ca_3PbH_{24}$ | - | $Y_3GeH_{24}$ | - |
| $Ca_3AuH_{24}$ | - | $Y_3CuH_{24}$ | - | $Y_3SnH_{24}$ | - |
| $Ca_3ZnH_{24}$ | - | $Y_3AgH_{24}$ | - | $Y_3PbH_{24}$ | - |
| $Ca_3CdH_{24}$ | - | $Y_3AuH_{24}$ | - | $Mg_3InH_{24}$ | - |
| $Ca_3HgH_{24}$ | - | $Y_3ZnH_{24}$ | - | $Mg_3CdH_{24}$ | - |
| $Ca_3GaH_{24}$ | - | $Y_3CdH_{24}$ | - | $La_3InH_{24}$ | - |
| $Ca_3InH_{24}$ | - | $Y_3HgH_{24}$ | - | $La_3CdH_{24}$ | - |
| $Ca_3TlH_{24}$ | - | $Y_3GaH_{24}$ | - | $Sc_3SnH_{24}$ | - |
| $Ca_3GeH_{24}$ | - | $Y_3InH_{24}$ | - | | |
| Dynamical stability of **$LaXH_8$** ($Fm\bar{3}m$) at 200 GPa | | | | | |
| $LaCuH_8$ | - | $LaCdH_8$ | - | $LaTlH_8$ | - |
| $LaAgH_8$ | - | $LaHgH_8$ | - | $LaGeH_8$ | - |
| $LaAuH_8$ | - | $LaGaH_8$ | - | $LaSnH_8$ | - |
| $LaZnH_8$ | - | $LaInH_8$ | - | $LaPbH_8$ | - |

Firstly, in order to check the dynamical stability, we calculated the phonon spectra of these predicted compounds at 200 GPa, as shown in Fig. S1-S14. As listed in Table 1 and Table 2, the negative sign indicates the presence of imaginary frequencies, and the positive sign indicates no imaginary frequencies along the Brillouin zone. We also retained structures with imaginary frequencies less than 20 $cm^{-1}$, which have the possibility to be stable within the increase of 50 GPa. After the first step, we screened out 3 $A_2XH_{18}$ ($P\bar{3}m1$) structures including $Y_2CdH_{18}$, $Y_2InH_{18}$ and $Ca_2SnH_{18}$, and 7 $AC_2H_8$ ($Fm\bar{3}m$) structures including $AgC_2H_8$, $AuC_2H_8$, $ZnC_2H_8$, $CdC_2H_8$, $HgC_2H_8$, $InC_2H_8$ and $TlC_2H_8$. As for $AXH_{20}$ ($P4/mmm$), $A_3XH_{24}$ ($Fm\bar{3}m$) and $LaXH_8$ ($Fm\bar{3}m$), we studied a total of 65 structures but none of them satisfy the first screening criterion. Secondly, we performed formation energy calculations after step one to study the energetic stability. Table 3 lists the values of the $AC_2H_8$ candidate structures, none of the structures have negative value, suggesting that they could decompose into simple substances at 200 GPa. Meanwhile, $Y_2CdH_{18}$, $Y_2InH_{18}$ and $Ca_2SnH_{18}$ exhibit negative formation energy under high pressure, which are potential for further thermodynamic stability screening. Thirdly, we conducted thermodynamic stability for candidate structures. Combined with phonon calculations, we identified their lowest dynamically stable pressures and calculated the relative enthalpy of $Y_2CdH_{18}$ at 250 GPa, $Y_2InH_{18}$ at 210 GPa and $Ca_2SnH_{18}$ at 180 GPa, respectively.

TABLE 3. The calculated formation energy of dynamically stable $AC_2H_8$ ($Fm\bar{3}m$) compounds at 200 GPa.

| | Space group | Pressure (GPa) | Formation Energy (eV/atom) |
|---|---|---|---|
| $AgC_2H_8$ | $Fm\bar{3}m$ ($NaC_2H_8$) | 200 | 0.21 |
| $AuC_2H_8$ | $Fm\bar{3}m$ ($NaC_2H_8$) | 200 | 0.32 |
| $ZnC_2H_8$ | $Fm\bar{3}m$ ($NaC_2H_8$) | 200 | 0.21 |
| $CdC_2H_8$ | $Fm\bar{3}m$ ($NaC_2H_8$) | 200 | 0.15 |
| $HgC_2H_8$ | $Fm\bar{3}m$ ($NaC_2H_8$) | 200 | 0.11 |
| $InC_2H_8$ | $Fm\bar{3}m$ ($NaC_2H_8$) | 200 | 0.06 |
| $TlC_2H_8$ | $Fm\bar{3}m$ ($NaC_2H_8$) | 200 | 0.21 |

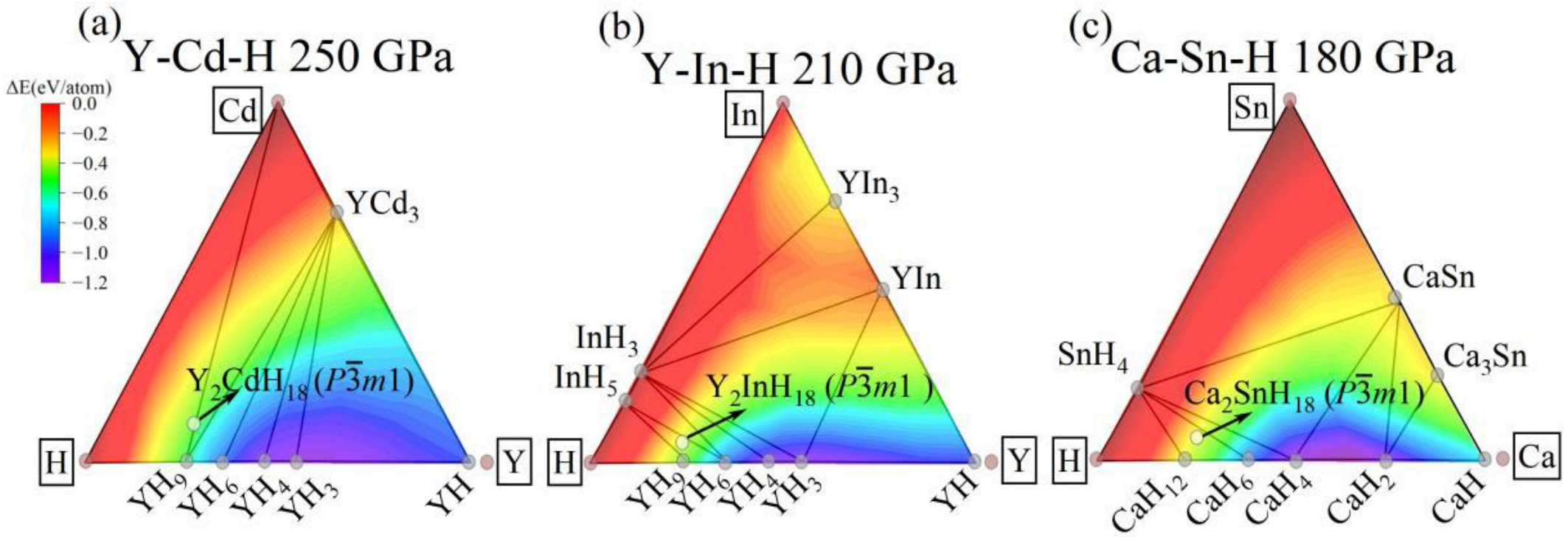

FIG. 2. The ternary convex hull of the (a) Y-Cd-H system at 250 GPa, (b)Y-In-H system at 210 GPa and (c) Ca-Sn-H system at 180 GPa. These solid gray dots indicate stable structures, the white dots indicate meta-stable structures.

According to previous reports [6, 7, 51-56], we constructed the relative enthalpy convex hull of Y-Cd-H system, Y-In-H system and Ca-Sn-H system at their corresponding pressure, as depicted in Fig. 2. Considering these compositions are in the hydrogen-rich region, the ternary convex hull starts from the A:H = 1:1 (A = Y, Ca) sites. To our knowledge, there are no stable Cd-H binary compounds under high pressure. Hence we performed variable-compositions structure search of Cd-H under 250 GPa, but we found no stable Cd-H compounds, suggesting that no compositions are on the Cd-H side of the ternary convex hull. As shown in Fig. 2, the predicted $Y_2CdH_{18}$, $Y_2InH_{18}$ and $Ca_2SnH_{18}$ are not on the ternary convex hulls. Thus, we further calculated the relative enthalpy difference between the convex hull and these predicted compositions, as shown in Table 4. In general, the criterion to distinguish meta-stability is 50 meV/atom above the convex hull [57]. As shown in Table 4, the relative enthalpy difference for both $Y_2InH_{18}$ at 210 GPa and $Ca_2SnH_{18}$ at 180 GPa are within the threshold for meta-stability. Nevertheless, the enthalpy difference of $Y_2CdH_{18}$ reaches 125.44 meV/atom at 250 GPa. Although this value is above 50 meV/atom, experimental reports in Ref.[58] synthesized some ternary metal compounds within the threshold of 200 meV/atom, and $Y_2CdH_{18}$ may exist in the quick annealing process such as laser heating under high pressure [59].

TABLE 4. The enthalpy difference relative to the convex hull for $Y_2CdH_{18}$, $Y_2InH_{18}$ and $Ca_2SnH_{18}$ under specific pressures.

| | Space group | Pressure (GPa) | Formation Energy (meV/atom) | Enthalpy difference (meV/atom) |
|---|---|---|---|---|
| $Y_2CdH_{18}$ | $P\bar{3}m1$ | 250 | -417.71 | 125.44 |
| | | 270 | -419.06 | 127.26 |
| $Y_2InH_{18}$ | $P\bar{3}m1$ | 210 | -457.71 | 42.56 |
| | | 230 | -461.98 | 40.26 |
| $Ca_2SnH_{18}$ | $P\bar{3}m1$ | 180 | -484.19 | 32.15 |
| | | 200 | -489.43 | 31.26 |

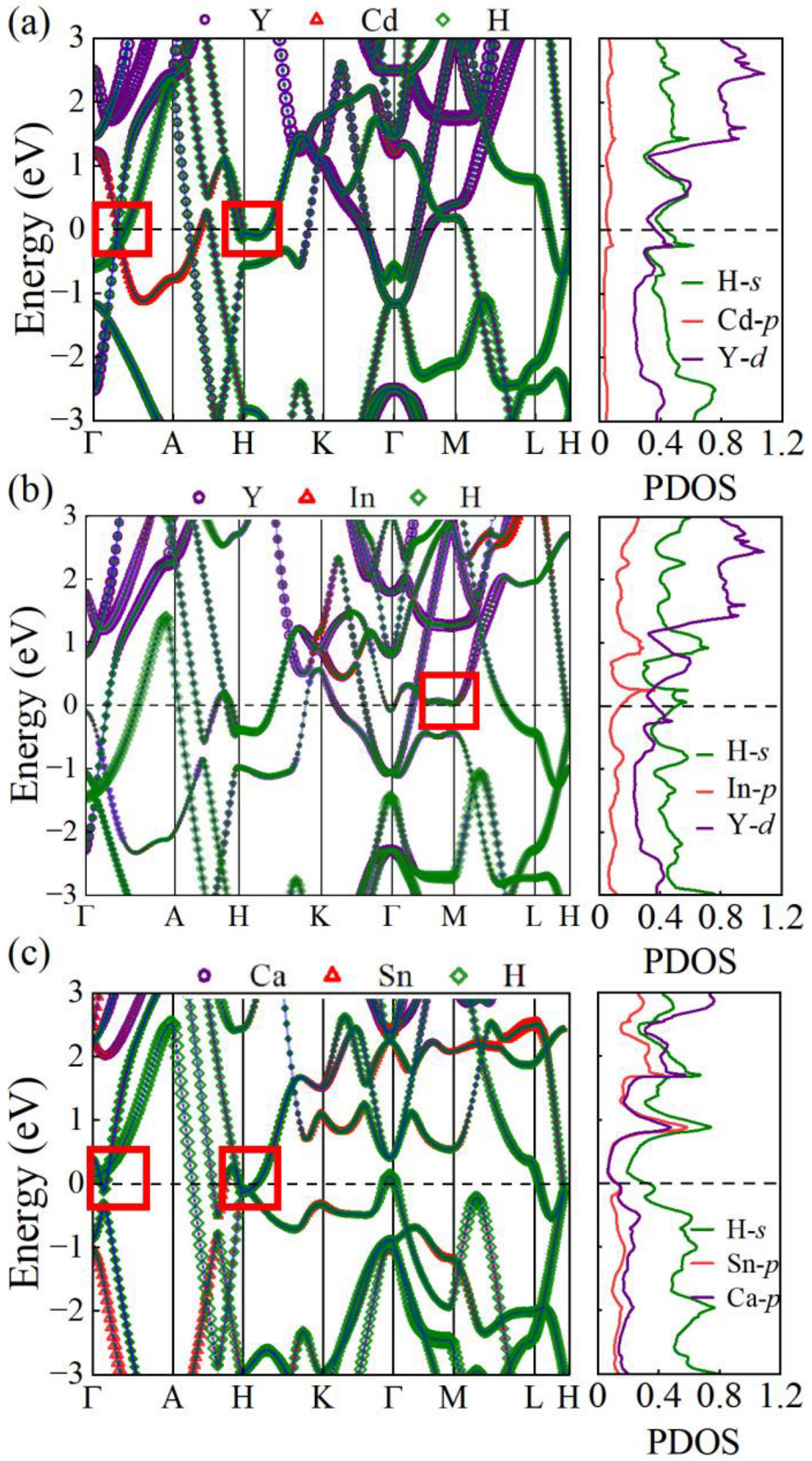

FIG. 3. The band structures and corresponding partial density of states of (a)$Y_2CdH_{18}$ at 250 GPa, (b) $Y_2InH_{18}$ at 210 GPa and (c) $Ca_2SnH_{18}$ at 180 GPa.

Then we calculated electronic band structures and partial density of states (PDOS) as illustrated in Fig. 3, from which we can find several bands crossing the Fermi energy ($E_F$), indicating metallic properties. In $Y_2CdH_{18}$ and $Y_2InH_{18}$, flat bands and typical crossings appear near the $E_F$. In $Ca_2SnH_{18}$, typical crossings appear along the Γ-A direction and around the H point. These band features contribute to the peaks in density of states (DOS) at $E_F$, which favors the emergence of superconductivity. In addition, the DOS at $E_F$ in $Ca_2SnH_{18}$ is dominated by hydrogen, while the contribution of Y-*d* electrons is comparable to that of H-*s* electrons in $Y_2CdH_{18}$ and $Y_2InH_{18}$, suggesting the DOS distribution from the metal elements are non-negligible.

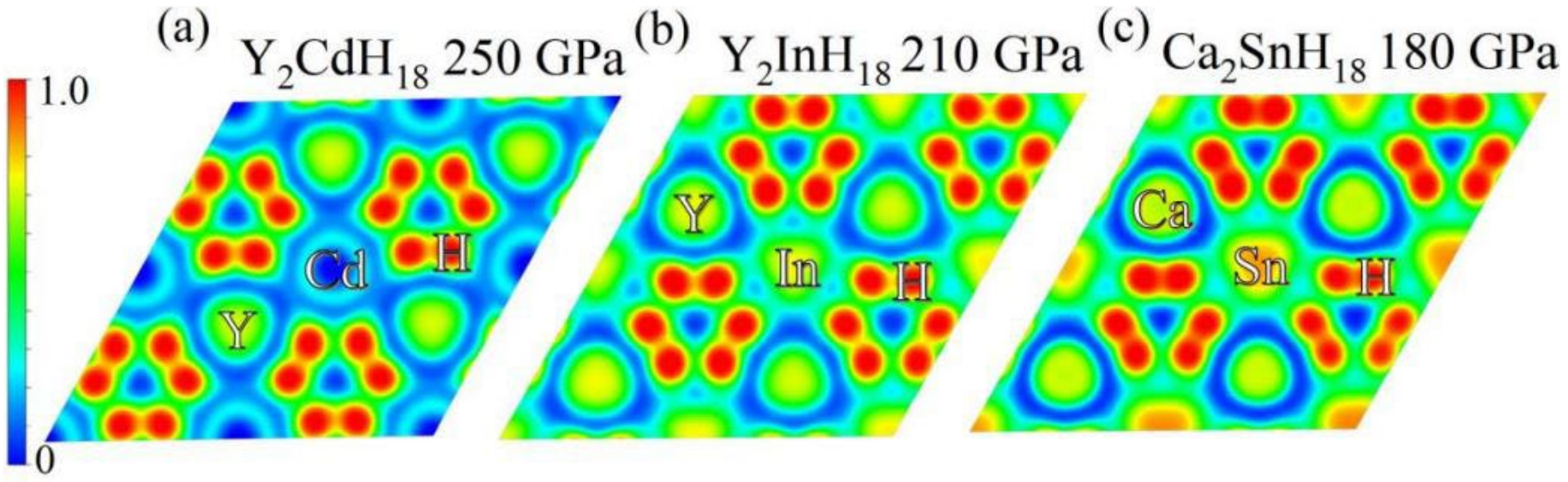

FIG. 4. Electronic localization function (ELF) on the (0, 0, 1) plane of (a) $Y_2CdH_{18}$ at 250 GPa and (b) $Y_2InH_{18}$ at 210 GPa and (c) $Ca_2SnH_{18}$ at 180 GPa.

To further explore the bonding condition of the predicted compounds under pressure, we conducted electron localization function (ELF) calculations. In Fig. 4(a), there are ionic bonding characteristics between Y-H and Cd-H, while there are charge distributions between In-H and Sn-H in Fig. 4(b) and 4(c). This illustrates the formation of electronic channels between In-H and Sn-H, indicating the characters of covalent bonds. In addition, we could observe the hexagonal H-H cage features in the three compounds, and the ELF suggest the presence of covalent bonding between H-H.

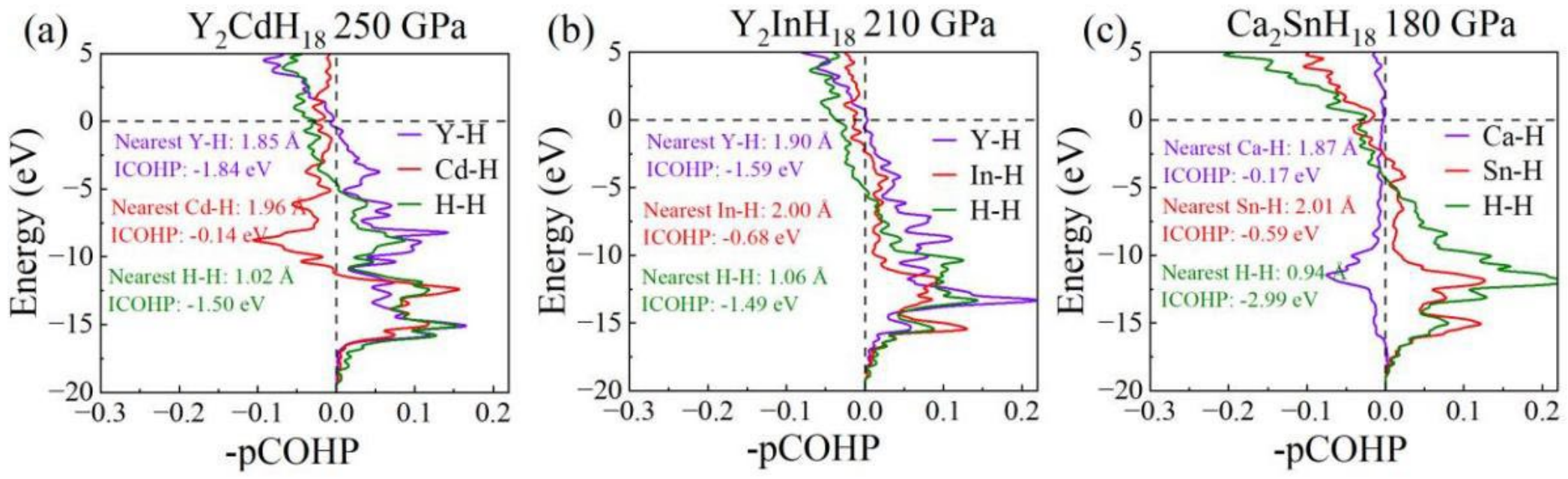

FIG. 5. The projected crystal orbital Hamilton populations (pCOHP) of the representative atom pairs in the predicted (a) $Y_2CdH_{18}$ at 250 GPa, (b) $Y_2InH_{18}$ at 210 GPa and (c) $Ca_2SnH_{18}$ at 180 GPa. The ICOHP results and corresponding bond lengths are listed in the inset. The Fermi energy is set to zero.

Then we conducted COHP calculations to determine the bond strength in these predicted compounds. According to the convention, the positive and negative values of pCOHP represent bonding and anti-bonding characteristics, respectively. We used the integral of COHP (ICOHP) to recognize the bonding strength and we calculated the integral of crystal orbital bond index (ICOBI) as well. As depicted in Fig. 5(a), Cd-H exhibits anti-bonding peak around -8.4 eV, while its overall ICOHP exhibits bonding characteristics. This leads to its ICOHP magnitude an order lower than that of Y-H. For $Y_2InH_{18}$, the main distribution of COHP is bonding. Comparing $Y_2InH_{18}$ and $Y_2CdH_{18}$ [Fig. 5(a) and 5(b)], their ICOHP values of Y-H and H-H are comparable, while the ICOHP value of In-H is about five times higher than that of Cd-H. This could attribute to the distinct bonding type between Cd-H and In-H. As for $Ca_2SnH_{18}$, Ca-H has a peak of anti-bonding at -11.6 eV but the ICOHP results indicate bonding characteristics. Analogous to $Y_2InH_{18}$ [Fig. 5(b)], the ICOHP value of Sn-H is around 3.5 times higher than that of Ca-H [Fig. 5(c)]. This illustrates the covalent type bonding in M-H (M=metal elements) has high strength than that of ionic type in $Y_2InH_{18}$ and $Ca_2SnH_{18}$. Moreover, the ICOHP value of the H-H bonds in $Ca_2SnH_{18}$ [Fig. 5(c)] is about twice than that of $Y_2CdH_{18}$ and $Y_2InH_{18}$ [Fig. 5(a) and 5(b)]. This may correspond to the relatively lower stable pressure of $Ca_2SnH_{18}$ among the three predicted compounds, and we assume that both the H-H bond strength and covalent type bonding in M-H (M=metal elements) may have influence on reducing the stable pressure in designing ternary hydrides. The calculated ICOBI values are listed in Table 5, which have consistent trend with ICOHP results.

TABLE 5. The calculated ICOBI of typical chemical bonds for $Y_2CdH_{18}$, $Y_2InH_{18}$ and $Ca_2SnH_{18}$ under specific pressures.

| | Space group | Pressure (GPa) | Y-H | Ca-H | Cd-H | In-H | Sn-H | H-H |
|---|---|---|---|---|---|---|---|---|
| $Y_2CdH_{18}$ | $P\bar{3}m1$ | 250 | 0.16 | -- | 0.01 | -- | -- | 0.12 |
| $Y_2InH_{18}$ | $P\bar{3}m1$ | 210 | 0.14 | -- | -- | 0.05 | -- | 0.12 |
| $Ca_2SnH_{18}$ | $P\bar{3}m1$ | 200 | -- | 0.01 | -- | -- | 0.06 | 0.32 |

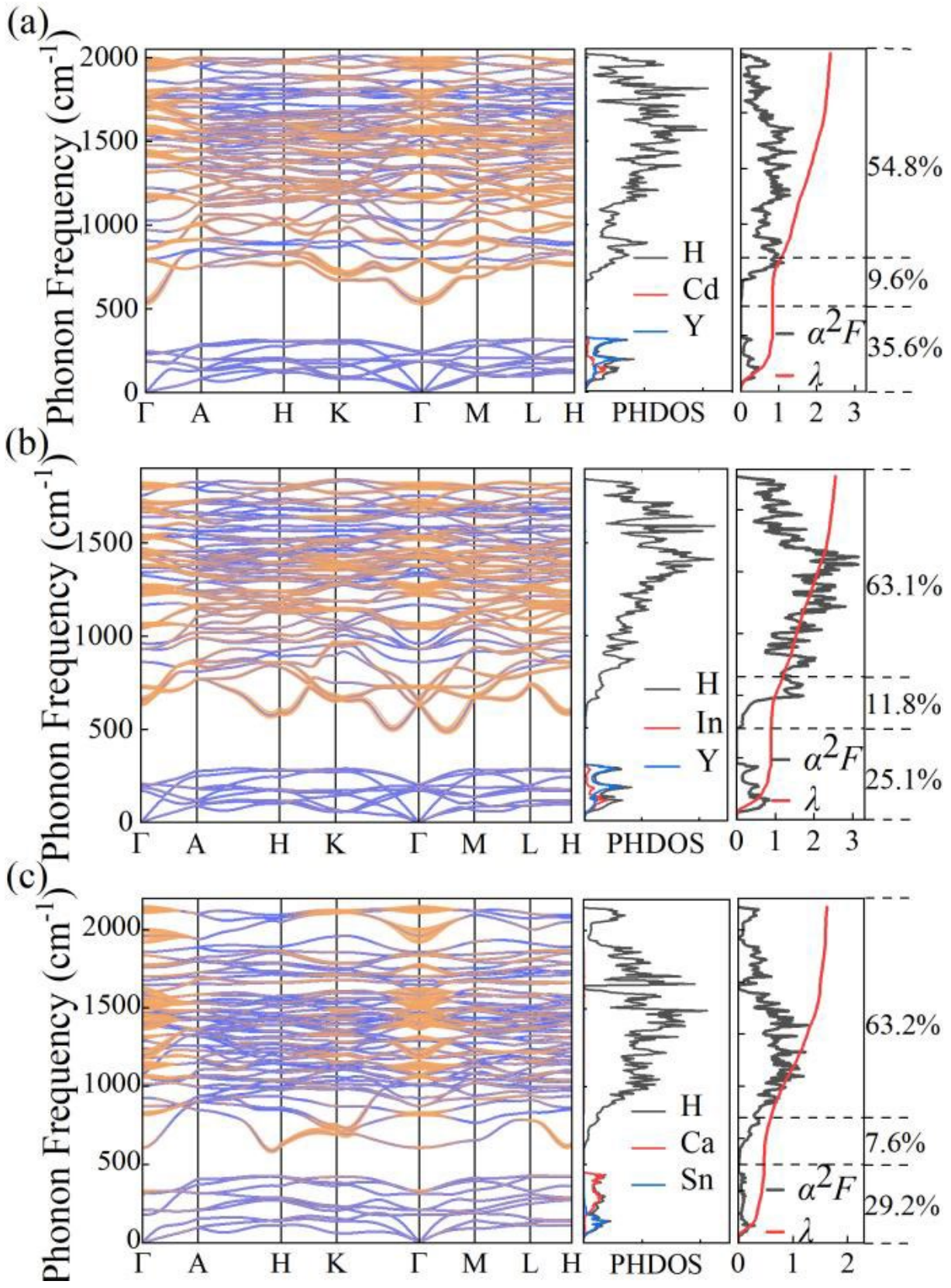

FIG. 6. The calculated phonon curves, projected phonon DOS, Eliashberg spectral functions $\alpha^2F$ and the EPC integral $\lambda$ of the predicted structures at their lowest dynamically stable pressure. The size of the orange solid dots represents the contribution to EPC.

We further calculated the phonon dispersion curves, projected phonon DOS (PHDOS), Eliashberg spectral functions $\alpha^2F(\omega)$ and the EPC integral $\lambda$ of $Y_2CdH_{18}$, $Y_2InH_{18}$ and $Ca_2SnH_{18}$, as shown in Fig. 6. The mode-resolved EPC constants $\lambda_{q,v}$ and the total EPC constant $\lambda$ can be expressed as [60]:

$$\lambda = \sum_{q,v} \lambda_{q,v}\, \omega_{q,v} \tag{4}$$

where $q,v$ are the phonon vector and the index of the mode, respectively, and $\omega_{q,v}$ is the weight parameter. We used the size of the orange dots to represent the contribution of each vibration modes to EPC, as plotted in Fig. 6. Combined with PHDOS, we divided phonon spectra into two regions with a boundary of 500 $cm^{-1}$. The phonon modes less than 500 $cm^{-1}$ are associated with metal elements Y, Ca, In, Cd and Sn, and the vibration modes of hydrogen atoms occupy the region above 500 $cm^{-1}$. The contribution of the H phonon modes in $Y_2CdH_{18}$, $Y_2InH_{18}$ and $Ca_2SnH_{18}$ is 64.4%, 74.9% and 70.8%, respectively, playing dominant roles in the EPC. Meanwhile, phonon modes related to metal elements corresponds to over 25% in the EPC integral $\lambda$ of the three compounds. This is in line with the PDOS distribution of metal atoms at $E_F$ in Fig. 3. Besides, we found some optical branches exhibiting soft phonon mode characteristics in the range of 500~700 cm$^{-1}$, and the contributions of these branches to EPC is 9.6%, 11.8% and 7.6%, respectively, which have significant impact to superconductivity as other hydrides under high pressure.

To estimate the superconducting $T_c$ of the predicted structures, we used Allen-Dynes modified McMillan formula. We set the Coulomb pseudopotential $\mu^*$ to 0.10 and 0.13, and the EPC constant $\lambda$ of $Y_2CdH_{18}$ at 250 GPa is 2.4 with the estimated $T_c$ 118 K. $Y_2InH_{18}$ has an estimated $\lambda$ value of 2.6 with the $T_c$ value of 113 K at 210 GPa. The value of $\lambda$ is 1.6 for $Ca_2SnH_{18}$ at 180 GPa, with $T_c$ reaching 111 K. The calculated EPC parameters and $T_c$ values of the three predicted hydrides are listed in Table 6, all of the estimated $T_c$ values exceed the liquid nitrogen temperature 77 K.

TABLE 6. The calculated EPC constant $\lambda$, logarithmic average phonon frequency $\omega_{log}$ (K), electronic DOS at fermi energy $N\varepsilon_f$ (states/eV/f.u.) of the predicted compounds. Two $T_c$ values were calculated with $\mu^*$ equal to 0.13 and 0.10, respectively.

| | Space group | Pressure (GPa) | $\lambda$ | $\omega_{log}$ | $N\varepsilon_f$ | $T_c$ (K) $\mu^*$=0.1 | $T_c$ (K) $\mu^*$=0.13 |
|---|---|---|---|---|---|---|---|
| $Y_2CdH_{18}$ | $P\bar{3}m1$ | 250 | 2.40 | 730.64 | 10.61 | 117.84 | 111.08 |
| $Y_2InH_{18}$ | $P\bar{3}m1$ | 210 | 2.60 | 670.95 | 10.91 | 112.81 | 106.71 |
| $Ca_2SnH_{18}$ | $P\bar{3}m1$ | 180 | 1.63 | 905.23 | 7.52 | 110.86 | 101.91 |

## Conclusion

In summary, we selected 5 compounds as prototype structures and adopted the elements substitution method to explore the potential superconductor under lower pressure. We utilized a three-step approach to screen and study 115 constructed structures after the substitution of 16 metal elements. Firstly, we studied their dynamical stability at 200 GPa. Secondly, we picked out candidate structures with negative formation energy. Thirdly, we performed thermodynamic stability research in the aspect of relative enthalpy and convex hull. We screened out three meta-stable structures including $Y_2CdH_{18}$, $Y_2InH_{18}$ and $Ca_2SnH_{18}$, all of which are superconductive with $T_c$ above the temperature of liquid nitrogen. Besides, we assume that both the H-H bond strength and covalent-type bond of M-H (M = metal elements) may help reducing the stable pressure in designing ternary hydrides. Our research investigate the effects of substitution of some unexpected elements in ternary hydrides and enriches the database of superconductor under high pressure, which may benefit to further theoretical and experimental researches in the future.

## Acknowledgment

This work was supported by the National Natural Science Foundation of China (Grant Nos. 52272265), and the National Key R&D Program of China (Grant No. 2023YFA1607400).

## References

[1] N.W. Ashcroft, Hydrogen Dominant Metallic Alloys: High Temperature Superconductors?, Phys. Rev. Lett. 92, 187002 (2004).

[2] D. Duan, Y. Liu, F. Tian, D. Li, X. Huang, Z. Zhao, H. Yu, B. Liu, W. Tian, T. Cui, Pressure-induced metallization of dense $(H_2S)_2H_2$ with high-$T_c$ superconductivity, Sci. Rep. 4, 6968 (2014).

[3] A.P. Drozdov, M.I. Eremets, I.A. Troyan, V. Ksenofontov, S.I. Shylin, Conventional superconductivity at 203 kelvin at high pressures in the sulfur hydride system, Nature. 73, 525 (2015).

[4] Y. Li, J. Hao, H. Liu, Y. Li, Y. Ma, The metallization and superconductivity of dense hydrogen sulfide, J. Chem. Phys. 140, 174712 (2014).

[5] H. Liu, I.I. Naumov, R. Hoffmann, N.W. Ashcroft, R.J. Hemley, Potential high-$T_c$ superconducting lanthanum and yttrium hydrides at high pressure, Proc. Natl. Acad. Sci. USA 114, 6990 (2017).

[6] F. Peng, Y. Sun, C.J. Pickard, R.J. Needs, Q. Wu, Y. Ma, Hydrogen Clathrate Structures in Rare Earth Hydrides at High Pressures: Possible Route to Room-Temperature Superconductivity, Phys. Rev. Lett. 119, 107001 (2017).

[7] H. Wang, J.S. Tse, K. Tanaka, T. Iitaka, Y. Ma, Superconductive sodalite-like clathrate calcium hydride at high pressures, Proc. Natl. Acad. Sci. USA 114, 6990 (2017).

[8] A.G. Kvashnin, D.V. Semenok, I.A. Kruglov, I.A. Wrona, A.R. Oganov, High-Temperature Superconductivity in a Th–H System under Pressure Conditions, ACS Appl. Mater. Interfaces 10, 43809(2018).

[9] X. Li, X. Huang, D. Duan, C.J. Pickard, D. Zhou, H. Xie, Q. Zhuang, Y. Huang, Q. Zhou, B. Liu, T. Cui, Polyhydride $CeH_9$ with an atomic-like hydrogen clathrate structure, Nat. Commun. 10, 3461 (2019).

[10] P. Kong, V.S. Minkov, M.A. Kuzovnikov, A.P. Drozdov, S.P. Besedin, S. Mozaffari, L. Balicas, F.F. Balakirev, V.B. Prakapenka, S. Chariton, D.A. Knyazev, E. Greenberg, M.I. Eremets, Superconductivity up to 243 K in the yttrium-hydrogen system under high pressure, Nat. Commun. 12, 5075 (2021).

[11] D. Duan, X. Huang, F. Tian, D. Li, H. Yu, Y. Liu, Y. Ma, B. Liu, T. Cui, Pressure-induced decomposition of solid hydrogen sulfide, Phys. Rev. B 91, 180502 (2015).

[12] M. Einaga, M. Sakata, T. Ishikawa, K. Shimizu, M.I. Eremets, A.P. Drozdov, I.A. Troyan, N. Hirao, Y. Ohishi, Crystal structure of the superconducting phase of sulfur hydride, Nat. Phys. 12, 835 (2016).

[13] F. Hong, L. Yang, P. Shan, P. Yang, Z. Liu, J. Sun, Y. Yin, X. Yu, J. Cheng, Z. Zhao, Superconductivity of Lanthanum Superhydride Investigated Using the Standard Four-Probe Configuration under High Pressures*, Chin. Phys. Lett. 37, 107401 (2020).

[14] L. Ma, K. Wang, Y. Xie, X. Yang, Y. Wang, M. Zhou, H. Liu, X. Yu, Y. Zhao, H. Wang, G. Liu, Y. Ma, High-Temperature Superconducting Phase in Clathrate Calcium Hydride $CaH_6$ up to 215 K at a Pressure of 172 GPa, Phys. Rev. Lett. 128, 167001 (2022).

[15] C.J. Pickard, R.J. Needs, Structure of phase III of solid hydrogen, Nat. Phys. 3, 473 (2007).
[16] L. Zhang, Y. Niu, Q. Li, T. Cui, Y. Wang, Y. Ma, Z. He, G. Zou, Ab initio prediction of superconductivity in molecular metallic hydrogen under high pressure, Solid State Commun. 141, 610 (2007).
[17] P. Cudazzo, G. Profeta, A. Sanna, A. Floris, A. Continenza, S. Massidda, E.K.U. Gross, Ab Initio Description of High-Temperature Superconductivity in Dense Molecular Hydrogen, Phys. Rev. Lett. 100, 257001 (2008).
[18] L. Boeri, R. Hennig, P. Hirschfeld, G. Profeta, A. Sanna, E. Zurek, W.E. Pickett, M. Amsler, R. Dias, M.I. Eremets, C. Heil, R.J. Hemley, H. Liu, Y. Ma, C. Pierleoni, A.N. Kolmogorov, N. Rybin, D. Novoselov, V. Anisimov, A.R. Oganov, C.J. Pickard, T. Bi, R. Arita, I. Errea, C. Pellegrini, R. Requist, E.K.U. Gross, E.R. Margine, S.R. Xie, Y. Quan, A. Hire, L. Fanfarillo, G.R. Stewart, J.J. Hamlin, V. Stanev, R.S. Gonnelli, E. Piatti, D. Romanin, D. Daghero, R. Valenti, The 2021 room-temperature superconductivity roadmap, J. Phys.: Condens. Matter 34, 183002 (2022).
[19] Y. Sun, J. Lv, Y. Xie, H. Liu, Y. Ma, Route to a Superconducting Phase above Room Temperature in Electron-Doped Hydride Compounds under High Pressure, Phys. Rev. Lett. 123, 097001 (2019).
[20] Z. Zhang, T. Cui, M.J. Hutcheon, A.M. Shipley, H. Song, M. Du, V.Z. Kresin, D. Duan, C.J. Pickard, Y. Yao, Design Principles for High-Temperature Superconductors with a Hydrogen-Based Alloy Backbone at Moderate Pressure, Phys. Rev. Lett. 128, 047001 (2022).
[21] M. Gao, X.-W. Yan, Z.-Y. Lu, T. Xiang, Phonon-mediated high-temperature superconductivity in the ternary borohydride $KB_2H_8$ under pressure near 12 GPa, Phys. Rev. B 104, L100504 (2021).
[22] X. Liang, A. Bergara, X. Wei, X. Song, L. Wang, R. Sun, H. Liu, R.J. Hemley, L. Wang, G. Gao, Y. Tian, Prediction of high-$T_c$ superconductivity in ternary lanthanum borohydrides, Phys. Rev. B. 104, 134501 (2021).
[23] D.V. Semenok, I.A. Troyan, A.G. Ivanova, A.G. Kvashnin, I.A. Kruglov, M. Hanfland, A.V. Sadakov, O.A. Sobolevskiy, K.S. Pervakov, I.S. Lyubutin, K.V.

Glazyrin, N. Giordano, D.N. Karimov, A.L. Vasiliev, R. Akashi, V.M. Pudalov, A.R. Oganov, Superconductivity at 253 K in lanthanum–yttrium ternary hydrides, Mater. Today 48, 18 (2021).

[24] H. Xie, D. Duan, Z. Shao, H. Song, Y. Wang, X. Xiao, D. Li, F. Tian, B. Liu, T. Cui, High-temperature superconductivity in ternary clathrate $YCaH_{12}$ under high pressures, J. Phys.: Condens. Matter 31, 245404 (2019).

[25] W. Zhao, D. Duan, M. Du, X. Yao, Z. Huo, Q. Jiang, T. Cui, Pressure-induced high-$T_c$ superconductivity in the ternary clathrate system Y-Ca-H, Phys. Rev. B 106, 014521 (2022).

[26] L. Liu, F. Peng, P. Song, X. Liu, L. Zhang, X. Huang, C. Niu, C. Liu, W. Zhang, Y. Jia, Z. Zhang, Generic rules for achieving room-temperature superconductivity in ternary hydrides with clathrate structures, Phys. Rev. B 107, L020504 (2023).

[27] M.-J. Jiang, Y.-L. Hai, H.-L. Tian, H.-B. Ding, Y.-J. Feng, C.-L. Yang, X.-J. Chen, G.-H. Zhong, High-temperature superconductivity below 100 GPa in ternary C-based hydride $MC_2H_8$ with molecular crystal characteristics (M= Na, K, Mg, Al, and Ga), Phys. Rev. B 105, 104511 (2022).

[28] P. Song, A.P. Durajski, Z. Hou, A. Ghaffar, R. Dahule, R. Szczęśniak, K. Hongo, R. Maezono, (La,Th)$H_{10}$: Potential High-$T_c$ (242 K) Superconductors Stabilized Thermodynamically below 200 GPa, J. Phys. Chem. C 128, 2656 (2024).

[29] L.-T. Shi, Y.-K. Wei, A.K. Liang, R. Turnbull, C. Cheng, X.-R. Chen, G.-F. Ji, Prediction of pressure-induced superconductivity in the novel ternary system $ScCaH2_n$ (n = 1–6), J. Mater. Chem. C 9, 7284 (2021).

[30] J. Wu, B. Zhu, C. Ding, C. Pei, Q. Wang, J. Sun, Y. Qi, Superconducting ternary hydrides in Ca-U-H under high pressure, J. Phys.: Condens. Matter 36, 165703 (2024).

[31] P. Song, Z. Hou, K. Nakano, K. Hongo, R. Maezono, Potential high-$T_c$ superconductivity in YCeHx and LaCeHx under pressure, Mater. Today Phys 28 100873 (2022).

[32] Y.X. Fan, B. Li, C. Zhu, J. Cheng, S.L. Liu, Z.X. Shi, Superconductive Sodalite-Like Clathrate Hydrides $MXH_{12}$ with Critical Temperatures of near 300 K under

Pressures, Phys. Status Solidi B 2400240 (2024).
[33] E. Orgaz, A. Aburto, Electronic Structure of Ternary Ruthenium-Based Hydrides, J. Phys. Chem. C 112, 15586 (2008).
[34] X. Wei, X. Hao, A. Bergara, E. Zurek, X. Liang, L. Wang, X. Song, P. Li, L. Wang, G. Gao, Y. Tian, Designing ternary superconducting hydrides with A15-type structure at moderate pressures, Mater. Today Phys 34 101086 (2023).
[35] A. Siddique, A. Khalil, B.S. Almutairi, M. Bilal Tahir, T. Ahsan, A. Hannan, H. Elhosiny Ali, H. Alrobei, M. Alzaid, Structural, electronic, mechanical and dynamical stability properties of $LiAH_3$ (A = Sc, Ti & V) perovskite-type hydrides: A first principle study, Chem. Phys. 568, 111851 (2023).
[36] M. Du, H. Huang, Z. Zhang, M. Wang, H. Song, D. Duan, T. Cui, High-Temperature Superconductivity in Perovskite Hydride Below 10 GPa, Adv. Sci. n/a, 2408370 (2024).
[37] Z. Li, X. He, C. Zhang, X. Wang, S. Zhang, Y. Jia, S. Feng, K. Lu, J. Zhao, J. Zhang, B. Min, Y. Long, R. Yu, L. Wang, M. Ye, Z. Zhang, V. Prakapenka, S. Chariton, P.A. Ginsberg, J. Bass, S. Yuan, H. Liu, C. Jin, Superconductivity above 200 K discovered in superhydrides of calcium, Nat. Commun. 13, 2863 (2022).
[38] G. Kresse, J. Furthmüller, Efficient iterative schemes for *ab initio* total-energy calculations using a plane-wave basis set, Phys. Rev. B 54, 11169 (1996).
[39] P. Hohenberg, W. Kohn, Inhomogeneous Electron Gas, Phys. Rev 136, B864 (1964).
[40] W. Kohn, L.J. Sham, Self-Consistent Equations Including Exchange and Correlation Effects, Phys. Rev 140, A1133 (1965).
[41] J.P. Perdew, K. Burke, M. Ernzerhof, Generalized Gradient Approximation Made Simple, Phys. Rev. Lett. 77, 3865 (1996).
[42] J.P. Perdew, J.A. Chevary, S.H. Vosko, K.A. Jackson, M.R. Pederson, D.J. Singh, C. Fiolhais, Atoms, molecules, solids, and surfaces: Applications of the generalized gradient approximation for exchange and correlation, Phys. Rev. B 46, 6671 (1992).
[43] P.E. Blöchl, Projector augmented-wave method, Phys. Rev. B 50, 17953 (1994).

[44] H.J. Monkhorst, J.D. Pack, Special points for Brillouin-zone integrations, Phys. Rev. B 13, 5188 (1976).

[45] A. Togo, I. Tanaka, First principles phonon calculations in materials science, Scr. Mater 108, 1 (2015).

[46] P.C. Müller, C. Ertural, J. Hempelmann, R. Dronskowski, Crystal Orbital Bond Index: Covalent Bond Orders in Solids, J. Phys. Chem. C 125, 7959 (2021).

[47] P. Giannozzi, S. Baroni, N. Bonini, M. Calandra, R. Car, C. Cavazzoni, D. Ceresoli, G.L. Chiarotti, M. Cococcioni, I. Dabo, A. Dal Corso, S. de Gironcoli, S. Fabris, G. Fratesi, R. Gebauer, U. Gerstmann, C. Gougoussis, A. Kokalj, M. Lazzeri, L. Martin-Samos, N. Marzari, F. Mauri, R. Mazzarello, S. Paolini, A. Pasquarello, L. Paulatto, C. Sbraccia, S. Scandolo, G. Sclauzero, A.P. Seitsonen, A. Smogunov, P. Umari, R.M. Wentzcovitch, QUANTUM ESPRESSO: a modular and open-source software project for quantum simulations of materials, J. Phys.: Condens. Matter 21, 395502 (2009).

[48] S. Baroni, S. de Gironcoli, A. Dal Corso, P. Giannozzi, Phonons and related crystal properties from density-functional perturbation theory, Rev. Mod. Phys 73, 515 (2001).

[49] P.B. Allen, R.C. Dynes, Transition temperature of strong-coupled superconductors reanalyzed, Phys. Rev. B 12, 905 (1975).

[50] G.M. Éliashberg, Interactions between electrons and lattice vibrations in a superconductor, Sov. Phys. -JEPT 11, 696 (1960).

[51] Y. Liu, D. Duan, F. Tian, H. Liu, C. Wang, X. Huang, D. Li, Y. Ma, B. Liu, T. Cui, Pressure-Induced Structures and Properties in Indium Hydrides, Inorg. Chem. 54, 9924 (2015.)

[52] M. Mahdi Davari Esfahani, Z. Wang, A.R. Oganov, H. Dong, Q. Zhu, S. Wang, M.S. Rakitin, X.-F. Zhou, Superconductivity of novel tin hydrides ($Sn_nH_m$) under pressure, Sci. Rep. 6, 22873 (2016).

[53] Z. Yang, D. Shi, B. Wen, R. Melnik, S. Yao, T. Li, First-principle studies of Ca–X (X=Si,Ge,Sn,Pb) intermetallic compounds, J. Solid. State. Chem 183, 136 (2010).

[54] G. Bruzzone, M.L. Fornasini, F. Merlo, Rare earth intermediate phases with cadmium, J. less-common. met. 30, 361 (1973).

[55] S.S. Chouhan, G. Pagare, M. Rajagopalan, S.P. Sanyal, First principles study of structural, electronic, elastic and thermal properties of YX (X = Cd, In, Au, Hg and Tl) intermetallics, Solid. State. Sci. 14, 1004 (2012).

[56] R. Sharma, G. Ahmed, Y. Sharma, Intermediate coupled superconductivity in yttrium intermetallics, Physica. C 540, 1 (2017).

[57] H. Xiao, Y. Dan, B. Suo, X. Chen, Comment on "Accelerated Discovery of New 8-Electron Half-Heusler Compounds as Promising Energy and Topological Quantum Materials", J. Phys. Chem. C 124, 2247 (2020).

[58] W. Sun, C.J. Bartel, E. Arca, S.R. Bauers, B. Matthews, B. Orvañanos, B.-R. Chen, M.F. Toney, L.T. Schelhas, W. Tumas, J. Tate, A. Zakutayev, S. Lany, A.M. Holder, G. Ceder, A map of the inorganic ternary metal nitrides, Nat. Mater 18, 732 (2019) .

[59] S. Di Cataldo, C. Heil, W. von der Linden, L. Boeri, $LaBeH_8$: Towards high-$T_c$ low-pressure superconductivity in ternary superhydrides, Phys. Rev B 104, L020511 (2021).

[60] Y. Ma, D. Duan, Z. Shao, H. Yu, H. Liu, F. Tian, X. Huang, D. Li, B. Liu, T. Cui, Divergent synthesis routes and superconductivity of ternary hydride $MgSiH_6$ at high pressure, Phys. Rev B 96, 144518 (2017).